\documentclass[showpacs,preprintnumbers,amsmath,amssymb,
eqsecnum,
twocolumn,tightenlines,prd
]{revtex4}

\usepackage{bm}

\usepackage{amsmath}

\usepackage{dcolumn}

\usepackage[dvips]{graphicx} 

\begin{document}

\bibliographystyle{apsrev}    
\title {Scattering of scalar particles by a black hole}
    
\author{M.Yu.Kuchiev$^{1}$} \email[Email:]{kuchiev@newt.phys.unsw.edu.au}
\author{V.V.Flambaum$^{1,2}$} \email[Email:]{flambaum@newt.phys.unsw.edu.au}
    
\affiliation{$^1$ School of Physics, University of New South Wales, Sydney
  2052, Australia}
\affiliation{$^2$ Institute for Advanced Study, Einstein drive,
Princeton, NJ 08540, USA
}

\date{\today}

    \begin{abstract}
      The absorption cross section for scalar particle impact on a
      Schwarzschild black hole is found. The process is dominated by
      two physical phenomena. One of them is the well-known greybody
      factor that arises from the energy-dependent potential barrier
      outside the horizon that filters the incoming and outgoing
      waves.  The other is related to the reflection of particles on
      the horizon (Kuchiev 2003). This latter effect strongly
      diminishes the cross section for low energies, forcing it to
      vanish in the infrared limit. It is argued that this is a
      general property, the absorption cross section vanishes in the
      infrared limit for scattering of particles of arbitrary spin.
    \end{abstract}
 
    
    \maketitle

    \section{Introduction}

    This work presents new qualitative features for scattering of
    particles by black holes.  Interest in the scattering problem was
    first inspired long ago by the discovery of the Penrose process
    \cite{penrose_69}, which allows an impact particle to bring energy
    out of the Kerr black hole. After the works of Zel'dovich
    \cite{zeldovich_71_72} and Misner \cite{misner_72} it became clear
    that the energy extraction from the Kerr black hole can be
    described in terms of superradiant scattering. The corresponding
    amplification factor was calculated numerically by Press and
    Teukolsky \cite{press_teukolsky_72,press_teukolsky_74} and
    analytically by Starobinsky \cite{starobinsky_73} for the scalar
    field and Starobinsky and Churilov \cite{starobinsky_churilov_73}
    for electromagnetic and gravitational waves. Independently, in
    parallel with this line of research, Unruh \cite{unruh_76} found
    the absorption cross section for scalar and fermion particles
    scattered off a Schwarzschild black hole.  The results of these
    and related works are well known, being summarized in books
    \cite{frolov_novikov_98,%
      chandrasekhar_92,fullerman_handler_matzner_88} that provide also
    further references on the subject.
    
    An important qualitative feature of the scattering problem is
    related to the well-known greybody factors that arise from
    energy-dependent potential barriers outside the horizon, which
    filter the incoming and outgoing waves producing a strong impact
    on the cross section. In particular, they make the absorption
    cross sections finite, proportional to the event horizon area in
    the infrared region \cite{area}.  The greybody factors also
    manifest themselves in the Hawking radiation process
    \cite{hawking_74,hawking_75} filtering the initially blackbody
    spectrum emanating from the horizon. Refs.\cite{grey_factor}
    describe a number of different aspects relevant to the greybody
    effect.  Similarly, the potential barriers manifest themselves in
    the effect of gravitational lensing (for theory and references see
    the book \cite{schneider_etal_92}) that, in particular, can be
    caused by strong bending of light in the vicinity of $r=(3/2)
    r_g>r_g$ \cite{virbhadra_ellis_00}.
    
    The necessity to take another look at the scattering problem is
    prompted by Refs. \cite{kuchiev_1_03,kuchiev_2_03,kuchiev_3_03}
    that claim that the horizon has a new unexpected property: it is
    able to partially reflect the incoming and outgoing waves.  In
    other words, the events that take place on the horizon mix the
    incoming and outgoing waves.  In simple physical terms this
    means that a particle approaching the vicinity of the horizon can
    bounce back, into the outside world.  We will call this property
    the {\it reflection from the horizon} (RH).  The effect, which is
    strong for low energy particles, has a purely quantum origin,
    classically the particle penetrates the horizon smoothly.
    
    The fact that the RH takes place strictly on the horizon
    distinguishes it from the greybody effect that happens outside of
    the horizon.  This difference becomes particularly prominent for
    low energies of the incoming particle, when the greybody effect
    manifests itself at very large distances, much larger than the
    radius of the horizon. Another distinctive feature of the RH is
    its universal nature. The probability that a particle is reflected
    from the horizon is governed by only two parameters, the Hawking
    temperature and the energy of the incoming particle \footnote{This
      is true for the Schwarzschild case, see details in
      \cite{kuchiev_3_03} that describes the RH for charged rotating
      black holes}, being independent of the particle spin. This is
    different from the greybody effect that strongly depends on the
    spin of a particle.  The fact that there appears the Hawking
    temperature is not coincidental, the RH can be used for
    alternative derivation of the Hawking radiation effect. (Reversing
    this argument, one can claim that the Hawking radiation supports
    the validity of the RH.)
    
    The existence of the RH should strongly manifest itself in
    scattering, making it necessary to reexamine the scattering
    abilities of black holes. This paper addresses this problem
    considering scattering of scalar particles by the Schwarzschild
    black hole.  The problem is solved analytically for low impact
    energies and numerically for arbitrary energies. The analysis
    presented reveals that the RH reduces the absorption cross
    section. In particular, it forces the absorption cross section to
    vanish in the infrared regime $\sigma_\mathrm{abs} \propto
    \varepsilon$ when $\varepsilon \ll \hbar c/r_g~m>0$. It is argued
    that this is a general feature, the absorption cross section
    vanishes in the infrared region for scattering of any particle off
    an arbitrary black hole.

\section{Modification of the $S$-matrix due
 to reflection from horizon}    

    Consider the scalar field $\phi(x)$ in the vicinity of the
    conventional Schwarzschild black hole with the metric
    \begin{equation}
      \label{schw}
      ds^2 = -\left(1-\frac{1}{r}\right)dt^2 + \frac{dr^2}{1-1/r}
    +r^2 d\Omega^2~,
    \end{equation}
    where $d \Omega^2 = d \theta ^2 + \sin^2 \theta d\varphi^2$.  The
    relativistic units $\hbar=c=1$ are used and supplemented by the
    condition $2GM=1$ on the gravitational constant $G$ and the mass
    of the black hole $M$, the Schwarzschild radius in these units
    reads $r_g \equiv 2GM = 1$.  The Klein-Gordon equation
    $-\partial_\mu ( \sqrt{-g} \,g^{\mu\nu} \partial_\nu \phi) =
    \sqrt{-g} \, m^2\,\phi$ for the field $\phi(x)$ in the
    Schwarzschild metric allows the separation of variables $\phi(x) =
    \exp(-i \varepsilon t)Y_{lm} (\theta,\varphi) \phi_{l}(r)$, where
    $\varepsilon,l,m$ are the energy, momentum and its projection,
    while $\phi_{l}(r)$ is a radial function that satisfies
\begin{eqnarray}
        \label{phi''}
&&\phi''_{l} 
+ \left(\frac{1}{r}+\frac{1}{r-1} \right) \phi'_l
\\ \nonumber && +
\left(   p^2 +  \frac{\varepsilon^2+
    p^2}{r-1}+\frac{\varepsilon^2}{(r-1)^2}
-\frac{ l(l+1) }{r(r-1)} \right) \phi_{l} = 0~.
      \end{eqnarray}
      Here $p$ is the momentum at infinity.  Let us describe with the
      help of Eq.(\ref{phi''}) the scattering of the scalar particle
      by the black hole.  At large radiuses $r\gg 1$ Eq.(\ref{phi''})
      reduces to a Coulomb-type equation with the effective Coulomb
      charge $Z=(\varepsilon^2+p^2)/2= \varepsilon^2(1+v^2)/2$, where
      $v$ is the velocity of the impact particle at infinity
      \footnote{There is also a modification of the momentum due to
        the term $\varepsilon^2/(r-1)^2$ in Eq.(\ref{phi''}),
        but one can neglect it in the low energy limit $\varepsilon
        \ll 1$.}.  Therefore at large distances the solution can be
      presented as
\begin{equation}\label{inf}
\phi_l (r) \rightarrow \frac{1}{r} \left(\,
A_l \exp(iz) + B_l \exp(-iz)\, \right) ~,
     \end{equation}
     where $z = pr +\nu \ln(2pr)+ \delta_l^{(C)}+l\pi/2$ and
     \begin{equation}
       \label{deltaC}
     \delta_l^{(C)} = \arg \,\Gamma(l+1-i\nu)
     \end{equation}
     is the Coulomb phase.  Here $\nu = Z/p =
     v \varepsilon(1+1/v^2)/2$ is the conventional Coulomb parameter.
     Clearly the two terms in Eq.(\ref{inf}) describe the incoming and
     outgoing waves.  The scattering properties can be expressed via
     the $S$-matrix that can be written as the ratio of the
     coefficients in front of the incoming and outgoing waves
     \cite{landau_lifshits_77} that, accordingly, equals
\begin{equation}\label{S} 
S_l  = (-1)^{l+1} \frac{A_l}{B_l} \exp \left(2i\delta_l^{(C)}\right)~.
    \end{equation}
    The event horizon $r=1$ is a regular singular point of
    Eq.(\ref{phi''}). In its vicinity the solution can be presented as
    $\phi_l \approx \exp( \mp i \varepsilon \ln(r-1)\,)$.  Here the
    waves with signs minus and plus describe the incoming and outgoing
    waves respectively. Since absorption of particles by the black
    hole is considered, it appeared appropriate to discard the
    outgoing wave, imposing the condition $\phi_l(r) \rightarrow
    \exp(\mp i \varepsilon \ln(r-1)\,)$ on the horizon $r \rightarrow
    1 $, see Ref. \cite{unruh_76}.  However, recent Refs.
    \cite{kuchiev_1_03,kuchiev_2_03,kuchiev_3_03} argue that this
    condition should be modified. The wave function describing the
    incoming particle should necessarily include an admixture of the
    outgoing wave
\begin{equation}\label{reflection}
\phi_l(r) \rightarrow \exp[-i \varepsilon \ln(r-1)\,]
+ {\cal R}\exp[\,i \varepsilon \ln(r-1)\,]~.
    \end{equation}
    These works interpreted this boundary condition as a statement
    that the event horizon is able to partially reflect particles.
    The absolute value of the reflection coefficient ${\cal R}$ found
    in \cite{kuchiev_1_03,kuchiev_2_03,kuchiev_3_03} reads
\begin{equation}\label{R}
|{\cal R} | = \exp[-\varepsilon/(2T)\,]~,
    \end{equation}
    where $T$ is the Hawking temperature, $T=1/4\pi$ for the
    Schwarzschild black hole. The origin and physical meaning of the
    RH were discussed in detail in the mentioned works.  Here only
    brief comments are appropriate.  Ref.\cite{kuchiev_1_03} derives
    Eq.(\ref{reflection}) from the general symmetry properties of the
    Schwarzschild geometry. The two disconnected areas that describe
    the outer region (areas I and III on the Kruskal plane
    \cite{kruskal_60}, see \cite{misner_thorne_wheeler_73}) are
    physically identical.  Therefore there is a discrete symmetry that
    relates values of the wave function in these two regions. It is
    shown in \cite{kuchiev_1_03} that the incoming wave by itself,
    i.e. the first term in Eq.(\ref{reflection}), cannot satisfy this
    symmetry condition, whereas a linear combination of the incoming
    and outgoing waves complies with it, provided the reflection
    coefficient obeys Eq.(\ref{R}). Another way to derive this result
    was suggested in \cite{kuchiev_2_03}. We will discuss and use it
    below in order to find the phase of the reflection coefficient
    (proving it to be zero for low energy particles).
    
    Let us find the relation between the $S$-matrix and the reflection
    coefficient ${\cal R}$. Since the latter decreases exponentially
    with energy we will consider first the low energy region
    $\varepsilon \ll 1$, where the RH is prominent, restricting our
    attention to the most important for this region case $l=0$ and
    denoting the s-wave as $\phi(r) \equiv \phi_0(r)$.  
    
    Following the approach of Ref. \cite{unruh_76} consider three
    regions of distances.  Region 1 we choose in the vicinity of the
    horizon $r \rightarrow 1$. The wave function here is given in
    Eq.(\ref{reflection}). As region 2 we take those ``intermediate''
    distances $1 < r < \infty$, where one can neglect the low energy
    and momentum in Eq.(\ref{phi''}). The solution of thus simplified
    equation can be written as
\begin{equation}\label{albe}
\phi(r) = \alpha \, \ln \frac{r-1}{r} + \beta~. 
    \end{equation}
    Compare now Eq.(\ref{reflection}) with (\ref{albe}) in the region
    of distances close, but not {\em very} close to the horizon
    $r-1\ll 1$, i.e.  the region where one can expand the incoming and
    outgoing waves in Eq.(\ref{reflection}) $\exp[\mp i\varepsilon
    \ln(r-1)\,] \simeq 1 \mp i\varepsilon \,\ln(r-1)$ and use
    simultaneously the asymptotic relation $\ln [\,(r-1)/r\,]\simeq
    \ln(r-1)$ in Eq.(\ref{albe}).  This procedure allows us to find
    the coefficients in Eq.(\ref{albe})
\begin{equation}\label{ab}
\alpha = - i \varepsilon (1-{ \cal R })
~, \quad \beta = 1+{ \cal R }~.
    \end{equation}
    Consider now region 3, the region of large separations $r\gg 1$.
    The wave function behavior here is governed by the effective
    Coulomb problem. Introducing the regular $F(r)$ and singular
    $G(r)$ solutions of the Coulomb problem, see
    \cite{landau_lifshits_77}, one can present the wave function here
    as their linear combination
\begin{equation}\label{FG}
\phi(r) = \frac{1}{r}( \,a \,F(r) + b \,G(r)\,)~.
    \end{equation}
    Taking $r$ {\em reasonably} large one makes Eqs.(\ref{albe}) and
    (\ref{FG}) valid simultaneously.  Eqs.(\ref{albe}),(\ref{ab}) give
    \begin{equation}\label{23}
\phi(r) \simeq  1+{ \cal R } + 
i \varepsilon (1-{ \cal R })\,\frac{1}{r}~.
    \end{equation}
    On the other hand, since $\varepsilon$ is low we can assume that $
    p r \ll 1 $ and use the asymptotic relations for small distances in
    the wave functions of the Coulomb problem
    \begin{equation}
      \label{FGsmall}
F(r) \simeq C\,pr~, \quad
    G\simeq 1/C, 
    \end{equation}
    see \cite{landau_lifshits_77}, where
\begin{equation} \label{c2}
C^2 = \frac{2\pi \nu}{ 1-\exp(-2 \pi \nu)} ~.
    \end{equation}
    Eq.(\ref{FG}) then gives
    \begin{equation}
      \label{asyFG}
\phi(r) \simeq a\,pC +\frac{b}{C\,r}~.
    \end{equation}
    Comparing Eqs.(\ref{23}),(\ref{asyFG}) we find the s-wave 
    coefficients in Eq.(\ref{FG})
    \begin{equation}\label{abfind}
      a = \frac{1+{\cal R} }{pC} ~,\quad b= i\varepsilon C
( 1-{\cal R} )
    \end{equation}
    Thus Eqs.(\ref{abfind}),(\ref{FG}) define the behavior of the wave
    function at large distances. In the asymptotic region
    $r\rightarrow \infty$ we can use the known formulas, see
    \cite{landau_lifshits_77}, for the Coulomb functions
    \begin{equation}
      \label{FGlarge}
      F(r) \simeq \sin \,z~, \quad G(r) \simeq \cos\,z~,
    \end{equation}
    where $z = pr + \nu \ln 2pr + \delta_0^{ ( \mathrm{C} )}$, that
    allow us to present Eq.(\ref{FG}) in an asymptotic form
    (\ref{inf}). As a result we find the coefficients $A_0,B_0$ in
    the latter
    \begin{eqnarray}
      \label{A0B0}
\!\!\!       A_0 &=& \frac{b-ia}{2i} = 
\frac{1}{2ipC}[\,1+{\cal R} - p\varepsilon C^2(1-{\cal R} )\,],
\\ \nonumber
\!\!\!      B_0 &=& \frac{b+ia}{2i} =  
\frac{-1}{2ipC}[\,1+{\cal R} + p\varepsilon C^2(1-{\cal R} )\,].
    \end{eqnarray}
    The corresponding $S$-matrix Eq.(\ref{S}) for the s-wave reads
\begin{equation}
   \label{S0}
  S_0  = \frac{ 1+{\cal R} - v\varepsilon^2 C^2(1-{\cal R} ) }
              {1+{\cal R} + v\varepsilon^2 C^2(1-{\cal R} ) }
\exp \left( \,2i \delta_0^{ ( \mathrm {C} ) } \, \right)~.
   \end{equation}
   The factor ${\cal R}$ in this formula arises from the RH. If one
   wishes to neglect this phenomenon, one can put ${\cal R}=0$ in
   Eq.(\ref{reflection}) and, correspondingly, in the $ S $-matrix
   (\ref{S0}).  Then Eq.(\ref{S0}) reproduces the results of
   \cite{unruh_76}, see discussion of Eq.(\ref{unruh}) below.  The
   factors $v\varepsilon^2 C^2$ in the $S $-matrix originate from the
   greybody effect, they are present even in the ${\cal R}=0$
   approximation.  They arise due to those events that take place at
   large separations $r \sim 1/Z = 2/[\,v^2 \varepsilon^2(1+1/v^2)\,] \gg
   1 $. This is in contrast with the RH, which happens strictly at the
   horizon $r=1$. Thus Eq.(\ref{S0}) accounts for both the greybody
   factor and the effect of the RH.

   Using conventional expression for the inelastic cross section, see
   e. g. \cite{landau_lifshits_77}, we find from Eq.(\ref{S0}) the
   absorption cross section in the s-wave that dominates the process
   in the low energy limit.
    \begin{eqnarray}
\label{abs1}
\sigma_\mathrm {abs} &=&\frac{\pi}{p^2} \left
      (\,1-|S_0|^2\,\right)=\frac{4\pi \xi C^2 } {v(1+
\xi v\varepsilon^2 C^2)^2}, 
\\  \label{xi}
\xi &=& \frac{1-R}{1+R}~.
     \end{eqnarray}

   \section{Reflection coefficient of the horizon }    
   
   The absolute value of the reflection coefficient ${\cal R}$ is
   given in Eq.(\ref{R}). In order to find its phase we follow the
   approach of Ref.\cite{kuchiev_3_03}.  Consider first the incoming
   wave in the outside region in the close vicinity of the horizon,
   $\phi_\mathrm{in}(r) = \exp[-i\varepsilon \ln(r-1)\,]$ for $r
   \rightarrow 1,~r>1$.  Continue it into the interior region $r<1$
   using an analytical continuation over the variable $r$ into the
   lower semiplane of the complex plane $r$.  This procedure allows
   one to find the incoming wave in the interior region in the
   vicinity of the horizon, $r\rightarrow 1,~r<1$,
   \begin{equation}
     \label{interior1}
   \phi_\mathrm{in}(r) = |{\cal R}|^{1/2}   \exp[ -i\varepsilon
   \ln(1-r)\,]~.
     \end{equation}
     It is suppressed compared to the outside region by a factor
     $|{\cal R}|^{1/2} = \exp(-\pi \varepsilon)$. Continue now the
     incoming wave further into the interior region $r<1$ using the
     differential equation (\ref{phi''}). In the vicinity of the
     origin $r\rightarrow 0$ the equation simplifies, its solution
     here can be presented as
   \begin{equation}
     \label{interior3}
   \phi_\mathrm{in}(r) = u \ln r + v~,
     \end{equation}
     where $u,v$ are constants defined below. We assume that the total
     wave function should satisfy the conventional regular condition
     $\phi(r) \rightarrow const$ at the origin. Since the incoming
     wave by itself exhibits a singular $\propto \ln r$ behavior,
     there should exist the outgoing wave that compensates the
     singularity in the vicinity of the origin. Thus we find the
     outgoing wave in the vicinity of the origin. Repeating now the
     arguments in the reverse order, we take this outgoing wave and
     continue it towards the horizon using the differential equation;
     then continue it over the horizon using the analytical
     continuation via the lower semiplane of the complex $r$-plane.
     As a result there appears the outgoing wave in the outside region
     described by the second term in Eq.(\ref{reflection}).
     Eq.(\ref{R}) follows from the fact that the horizon is crossed
     twice, first by the incoming wave and then by the outgoing wave.
     Each crossing gives the suppression factor $|{\cal R}|^{1/2}$
     that combine to produce $|{\cal R}|$ in (\ref{R}). The method of
     crossing of the horizon used in this derivation relies on the
     analytical properties of the wave function that are related to
     the fundamental causality principle, which makes the method
     reliable.  
     
     Consider now the incoming wave in the inside region. Its behavior
     in the vicinity of the horizon and at the origin is defined by
     Eqs.(\ref{interior1}) and (\ref{interior3}). Take the
     ``intermediate'' region $0<r<1$, defined in such a way that
     inside this region one can neglect the small terms proportional
     to $\varepsilon,p$ in Eq.(\ref{phi''}). The solution in this
     region reads
     \begin{equation}
       \label{intRegion}
       \phi_\mathrm{in}(r) = \alpha ' \ln\frac{1-r}{r} +\beta '~.
     \end{equation}
     Consider now the region close, but not {\em very} close to the
     horizon $r\approx 1$, where Eqs.(\ref{interior1}) and
     (\ref{intRegion}) are both valid. Expand Eq.(\ref{interior1}) in
     powers of $\varepsilon \ln(1-r)$, $\phi (r) \simeq |{\cal
       R}|^{1/2} [\, 1-i\varepsilon \ln(r-1)\,]$ and rewrite
     Eq.(\ref{intRegion}) for this region as $\alpha '\ln (1-r) +\beta
     '$. This procedure gives the coefficients in Eq.(\ref{intRegion})
     \begin{equation}
       \label{a'b'}
\alpha'= -i
     |{\cal R}|^{1/2}\varepsilon, \quad \beta' = |{\cal R}|^{1/2}~.
     \end{equation}
     Consider now the region close, but not {\em very} close to the
     origin $r\approx 0$, where Eqs.(\ref{interior3}) and
     (\ref{intRegion}) are valid simultaneously. Rewriting
     (\ref{intRegion}) in this region as $\phi_\mathrm{in}(r) = -
     \alpha ' \ln{r} +\beta '$, we find the coefficients
     \begin{eqnarray}
       \label{uv}
       u = -\alpha'= i |{\cal R}|^{1/2}\varepsilon~, \quad
%\\ \nonumber~
       v  = ~~\beta' =~ |{\cal R}|^{1/2}~,
     \end{eqnarray}
     which define the behavior of the incoming wave at the origin in
     Eq.(\ref{interior3}). It is important that the coefficient $u$ in
     front of the singular term in Eq.(\ref{interior3}) is imaginary,
     while $v$ that describes the regular part of the solution is
     real. To see consequences of this fact let us take the outgoing
     wave $\phi_\mathrm{out}(r)$ that behaves at the horizon
     $r\rightarrow 1$ as
     \begin{equation}
       \label{out}
   \phi_\mathrm{out}(r) = |{\cal R}|^{1/2}   \exp[ \,i\varepsilon
   \ln(1-r)\,]~.
     \end{equation}
     According to Eq.(\ref{interior1}) this outgoing wave is simply a
     complex conjugate of the incoming wave. From Eq.(\ref{interior3})
     we find that $ \phi_\mathrm{out}(r) \rightarrow  u^*\ln r + v^*$
     when $r\rightarrow 0$. Using Eq.(\ref{uv}) we conclude that the
     wave function defined as
     \begin{equation}\label{InOut}
\phi(r) =
   \phi_\mathrm{in}(r)+ \phi_\mathrm{out}(r)
     \end{equation}
     is regular at the origin, $\phi(r)\rightarrow 2 v = |{\cal
       R}|^{1/2}$ when $r \rightarrow 0$. Using the analytical
     continuation described above for crossing of the event horizon,
     we find that in the outside region $r>1$ the wave function
     defined in Eq.(\ref{InOut}) coincides with the wave function in
     Eq.(\ref{R}). The fact that the coefficients in front of the
     incoming and outgoing waves in Eq.(\ref{InOut}) are identical
     (unity) permits one to find the reflection coefficient
     \cite{kuchiev_2_03}
       \begin{equation}
         \label{RR}
     {\cal R} = \exp[\,-\varepsilon/(2T) \,] = \exp(-2 \pi \varepsilon)~.
     \end{equation}
     The absolute value of the reflection coefficient found here
     agrees with Eq.(\ref{R}) (though Eq.(\ref{RR}) is derived in the
     low energy limit $\varepsilon \ll 1$, while Eq.(\ref{R}) remains
     valid for arbitrary energies; compare
     \cite{kuchiev_1_03,kuchiev_3_03} for alternative methods of
     derivation of Eq.(\ref{R})).

     \section{Absorption cross-section}    
     
     We can now substitute the reflection coefficient from (\ref{RR})
     into the $S$-matrix in Eq.(\ref{S0}). Expanding where possible
     the result in powers of $\varepsilon$ we find
     \begin{eqnarray}
       \label{Sfinal}
  S_0  &=& \frac{1-  v\varepsilon^2 C^2 \tanh (\pi\varepsilon) }
{1 + v \varepsilon^2 C^2 \tanh (\pi \varepsilon) }   
   \exp \left( \,2i \delta_0^{ ( \mathrm {C} ) } \, \right) 
\\ \label{SfinalSimple}
&\simeq&
( \,1 - 2 \pi v\varepsilon^3 C^2\,)
   \exp \left( \,2i \delta_0^{ ( \mathrm {C} ) } \, \right)~.
     \end{eqnarray}
     Correspondingly, Eq.(\ref{abs1}) gives the absorption cross
     section
    \begin{eqnarray}
\label{sigmaAnalytical}
\sigma_\mathrm {abs} &=&   
\frac{ 4 \pi C^2 \tanh (\pi \varepsilon) }
{v\Big(\,1+v \varepsilon^2 C^2\tanh (\pi \varepsilon) \,\Big)^2 } 
\\ \label{final}
&\simeq& \frac{4\pi^2 \varepsilon C^2}{v}  
= \frac{ 4\pi^3 \varepsilon^2( 1+1/v^2)  }
{1-\exp[\,-\pi v \varepsilon(1+1/v^2)\,] },
    \end{eqnarray}
    where Eq.(\ref{c2}) was used.  The last equation is presented for
    $ \pi \varepsilon \ll 1$.  For massless particles (or any particle
    with $ \pi \varepsilon/v \ll 1$)
\begin{equation}
  \label{vanish}
      \sigma_\mathrm {abs} \simeq 4\pi^2 \varepsilon \equiv 
\frac{4\pi^2 \varepsilon r_g^3}{\hbar c}  ~, 
    \end{equation}
    (the last expression in absolute units) the cross section vanishes
    in the infrared limit $\varepsilon \rightarrow 0$. In contrast,
    for slow massive particles $v\rightarrow 0$ ($ \varepsilon \ll 1$,
    $ \varepsilon/v \gg 1$) the cross section diverges as
\begin{equation}
  \label{diverg}
      \sigma_\mathrm {abs} = 
\frac{4\pi^3 m^2}{ v^2}
\equiv \frac{4\pi^3 m^2 c^4 r_g^4 }{\hbar^2 v^2}~,
    \end{equation}
    (absolute units). Compare these results with the previously known
    ones. If one neglects the RH effect, putting ${\cal R} =0$ in the
    $S$-matrix (\ref{S}) and substituting the latter into
    Eq.(\ref{abs1}) then one finds the cross section
      \begin{equation}
        \label{unruh}
       \sigma_\mathrm{abs}^{( {\cal R}=0) } = 
\frac{ 4\pi^2 \varepsilon( 1+1/v^2)  }
{1-\exp[\,-\pi v \varepsilon(1+1/v^2)\,] }~,
     \end{equation}
     that was first derived by Unruh \cite{unruh_76}.  Comparing
     Eqs.(\ref{final}),(\ref{unruh}) we see that for low energy
     particles the effect of the RH reduces the cross section by a
     factor of $\pi \varepsilon \equiv \pi \varepsilon r_g/\hbar c \ll
     1$, which is particularly important for massless particles in the
     infrared region where Eq.(\ref{unruh}) predicts the constant
     cross section
     \begin{equation}
       \label{unruhIR}
     \sigma_\mathrm{abs} = 4 \pi r_g^2~,
     \end{equation}
     (absolute units), which differs qualitatively from
     Eq.(\ref{vanish}) that describes the vanishing cross section. Let
     us repeat that the arguments of
     \cite{kuchiev_1_03,kuchiev_2_03,kuchiev_3_03}, some of which are
     partially reproduced above, indicate that the horizon possesses
     the reflective property  that inevitably leads to
     Eq.(\ref{final}).
\noindent
\begin{figure}[tbh]
  \centering \includegraphics[ height=5.5cm, keepaspectratio=true]{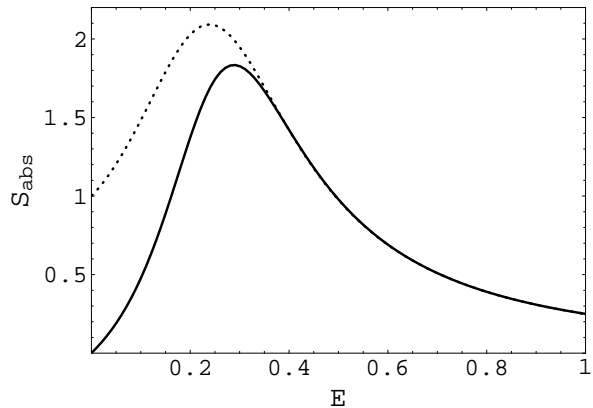}
\caption{The absorption cross section $S_\mathrm{abs}$
  (in units of the horizon area, $S_\mathrm{abs} \equiv
  \sigma_\mathrm{abs}/(4\pi r_g^2$)) for the s-wave scattering ($l=0$)
  of massless scalar particles by the Schwarzschild black hole versus
  the energy of the particle $E$ (dimensional units $E \equiv
  \varepsilon r_g/\hbar c$).  Solid line - the reflection on the
  horizon is taken into account, for low energy
  Eqs.(\ref{final}),(\ref{vanish}) of this work are valid.  Dashed
  line - the reflection by the horizon is not taken into account, at
  low energy this cross section agrees with
  Eqs.(\ref{unruh}),(\ref{unruhIR}) derived by Unruh \cite{unruh_76}.
}\label{one}
   \end{figure}
   \noindent

   The approach described above was also implemented numerically.  The
   differential equation (\ref{phi''}) was solved separately in the
   region $0< r < \infty$, using the discussed analytical continuation
   over the event horizon to match solutions in the inside and outside
   areas.  (The regular at the origin solution is presented as a
   linear combination of the incoming and outgoing waves, each wave is
   continued over the horizon using the analytical continuation into
   the lower semiplane of the complex $r$-plane.) The numerical method
   permits one to find the cross section for arbitrary, not
   necessarily low energies.  Generally speaking, such analysis needs
   that higher multipoles of the wave function be included.  However,
   for energies $\varepsilon \le 1$ the s-wave alone should give
   reasonable results.  Fig.  \ref{one} presents our results for the
   s-wave absorption cross section of massless scalar particles.  For
   $\varepsilon \ll 1$ they reproduce Eq.(\ref{final}). Fig.
   \ref{one} presents also results of similar calculations for the
   cross section when one neglects the reflective ability of the
   horizon, putting ${\cal R}=0$ in Eq.  (\ref{reflection}) (which
   allows one to formulate the problem entirely in the outside
   region).  Fig.  \ref{one} shows that the reflective ability of the
   horizon strongly diminishes the cross section in the low-energy
   region producing smaller impact for higher energies. This fact
   agrees qualitatively with Eq.(\ref{R}) that states that the
   reflection coefficient diminishes exponentially with the energy
   increase.

    \section{Discussion}
    As is known a classical particle can not reach the horizon during
    a finite interval of time in the reference frame of the external
    observer described by the metric Eq.(\ref{schw}). In this
    (restrictive) sense, the horizon represents an impenetrable
    barrier for the incoming particle. In the quantum process a role
    of the horizon is more subtle.  One can look at it considering the
    conserving current corresponding to the Klein-Gordon equation
    $-\partial_\mu ( \sqrt{-g} \,g^{\mu\nu} \partial_\nu \phi) =
    \sqrt{-g} \, m^2\,\phi$:
\begin{equation}
\label{curent}
j^\mu = \frac{1}{2i} \sqrt{-g} \, g^{\mu\nu} ( \phi^* \partial_\nu \phi -
 \phi \partial_\nu \phi^* )
   \end{equation}
   Near the horizon for $r>1$ the incoming wave function
   $\phi_\mathrm{in}(r)=\exp[-i \varepsilon \ln(r-1)\,]$ gives the
   following current components
\begin{eqnarray}
\label{j0}
j^0 &=& ~~\varepsilon\, \frac{r}{r-1} \,\sqrt{-g}~,\\ 
\label{jr}
j^r& =& - \varepsilon \, \frac{1}{r}\,\sqrt{-g}~.
   \end{eqnarray}
   We see that for the considered stationary solution the radial
   current does not have any singularity at the horizon $r=1$.  Zero
   velocity of the particle (in the reference frame of the external
   observer) is compensated by the infinite density.
   
   Consider now the current near horizon inside the black hole for the
   incoming wave $\phi_\mathrm{in}(r)= \sqrt{ |{\cal R}| } \,\exp[-i
   \varepsilon \ln(1-r)\,]$
\begin{eqnarray}
\label{j0i}
j^0 &=& -\varepsilon \,|{\cal R}| \,\frac{r}{1-r} \,\sqrt{-g}~, \\ \
\label{jri}
j^r &=&- \varepsilon\,|{ \cal R}|\, \frac{1}{r}\,\sqrt{-g} ~.
   \end{eqnarray}
   The radial current is still directed towards the center.  However,
   the zero-th component, i. e. the probability density, becomes
   negative.  A possible interpretation of this result can be related
   to a state of the ``hole'' that is produced inside. We call here by
   the hole a negative-energy state of the scalar field. Another
   popular way to call this state is to dub it as the ``antiparticle
   with negative energy''. (The {\it hole} considered here should not,
   of course, be confused with the {\it black hole} itself.)
   Propagation of the hole from the origin $r=0$ to the horizon
   creates the radial current towards the origin.  Similarly we can
   consider the outgoing wave inside $\phi_\mathrm{out}(r)=\sqrt{
     |{\cal R} | } \, \exp[i \varepsilon \ln(1-r)\,]$, which describes
   the current directed from the origin to the horizon.  In this wave
   the hole is produced on the horizon $r=1$ and moves towards the
   origin $r=0$.  The fact that the wave function inside
   Eq.(\ref{InOut}) includes both the incoming and outgoing waves
   shows that the state of the hole inside produces zero radial
   current.  To finish the argument, the outgoing wave in the outside
   region $\phi_\mathrm{out}(r)=|{ \cal R} | \exp[\,i \varepsilon
   \ln(r-1)\,]$ obviously describes the outgoing particle that moves
   from the horizon $r=1$ to infinity.

   Combining all peaces together, one can say that the horizon is
   responsible for the creation of a particle-hole pair and for the
   annihilation of another particle-hole pair.  It creates the
   incoming hole that moves inside and the outgoing particle that
   moves from the horizon to infinity. The hole created on the horizon
   moves towards the origin, is reflected there back towards the
   horizon, where its encounter with the incoming particle results in
   their mutual annihilation. The net result is the outgoing particle
   outside, which is exactly the RH.
    
   Here we find again a certain similarity between the Hawking
   radiation and the RH. In both phenomena the horizon creates the
   particle-hole pair with the probability that depends exponentially
   on $\varepsilon/T$. There is also a distinction. In the RH
   phenomenon a creation of one pair is accompanied by annihilation of
   another pair on the horizon, which shows that it is a more
   complicated phenomenon compared to the Hawking radiation.

\section{Conclusion}

   We discussed above two effects that combine together to make the
   absorption cross section small at low energies.  The first one is
   the known greybody effect that takes place outside of the event
   horizon, at the radius of the Coulomb zone that is large, $r\gg
   r_g$, for low energies. In the Coulomb zone the semiclassical
   approximation is violated making possible the reflection of the
   incoming wave. This reflection reduces the absorption cross
   section, making it finite, proportional to the area of the horizon
   in the infrared limit.
   
   Another opportunity for the reflection is presented by the RH
   effect that takes place strictly on the horizon. It reduces the
   cross section further, forcing it to vanish in the infrared limit.
   Importantly, one can expect that this is a general property of the
   cross section valid for the scattering of particles of arbitrary
   spins. This expectation is based on the fact that Refs.
   \cite{kuchiev_1_03,kuchiev_2_03,kuchiev_3_03} appeal to the
   semiclassical nature of the radial wave function in the vicinity of
   the horizon, being valid for the scattering of particles of
   arbitrary spins. There is a subtlety here, to verify this
   expectation one needs to prove that the phase of the reflection
   coefficient turns either $0$, or $\pi$ (which makes $\xi$ in
   Eq.(\ref{xi}) either $0$, or $\infty$), in the infrared region for
   arbitrary spins. This is a very plausible opportunity, but,
   strictly speaking, the present work does not discuss it.  It is
   also plausible that a similar more general statement can be
   formulated for scattering of particles of any charge and spin off a
   general Kerr-Newman black hole. The reflection coefficient in this
   case \cite{kuchiev_3_03} can be large, $| {\cal R}| =1$ for
   particular values of energy, charge and spin of the impact
   particle, which makes the reflection perfect and absorption
   impossible (though, again, to prove this result the particular
   phase conditions should be verified).  Several last remarks provide
   possible ways for further study.
   
   In conclusion, it is shown that the reflection on the horizon
   strongly influences the absorption cross section of a black hole,
   forcing it to vanish in the infrared limit.
     
   This work was supported by the Australian Research Council.  V.F.
   is grateful to the Institute for Advanced Study and Monell
   foundation for hospitality and support.

\end{document}